\documentclass[letter]{aa}  
\usepackage{natbib}
\def\kms{\hbox{km$\;$s$^{-1}$}}
\def\halpha{H$\mathrm{\alpha}$}
\def\cak{\hbox{\ion{Ca}{ii}~K}}
\def\Mgk{\hbox{\ion{Mg}{ii}~k}}
\usepackage{graphicx}
\usepackage{epstopdf}
\usepackage{txfonts}
\usepackage{hyperref}
\begin{document} 
  \title{Characterization and formation of on-disk spicules in the \cak{} and \Mgk{} spectral lines}
  \subtitle{}
  \author{Souvik Bose\inst{1}$^,$\inst{2}
    \and
    Vasco M.J. Henriques\inst{1}$^,$\inst{2}
    \and
    Jayant Joshi\inst{1}$^,$\inst{2}
    \and
    Luc Rouppe van der Voort\inst{1}$^,$\inst{2}
          }

  \institute{Institute of Theoretical Astrophysics, University of Oslo, P.O. Box 1029 Blindern, NO-0315 Oslo, Norway
          \and
    Rosseland Centre for Solar Physics, University of Oslo, P.O. Box 1029 Blindern, NO-0315 Oslo, Norway\\
             \email{souvik.bose@astro.uio.no}
             }

  \date{Received 06.09.2019; accepted 11.10.2019}

  \abstract
   {We characterize, for the first time, type-II spicules in \ion{Ca}{ii}~K 3934\AA\ using the CHROMIS instrument at the Swedish 1-m Solar Telescope. We find that their line formation is dominated by opacity shifts with the K$_{3}$ minimum best representing the velocity of the spicules. The K$_{2}$ features are either suppressed by the Doppler-shifted K$_{3}$ or enhanced via increased contribution from the lower layers, leading to strongly enhanced but un-shifted K$_{2}$ peaks, with widening towards the line-core as consistent with upper-layer opacity removal via Doppler-shift. We identify spicule spectra in concurrent IRIS \ion{Mg}{ii}~k~2796\AA\ observations with very similar properties. Using our interpretation of spicule chromospheric line-formation, we produce synthetic profiles that match observations.
  
   }

  \authorrunning{Bose et al.}

\maketitle


\section{Introduction}
\label{Section:intro}

\begin{figure*}
   \centering
   \includegraphics[width=\textwidth,height=16.5cm]{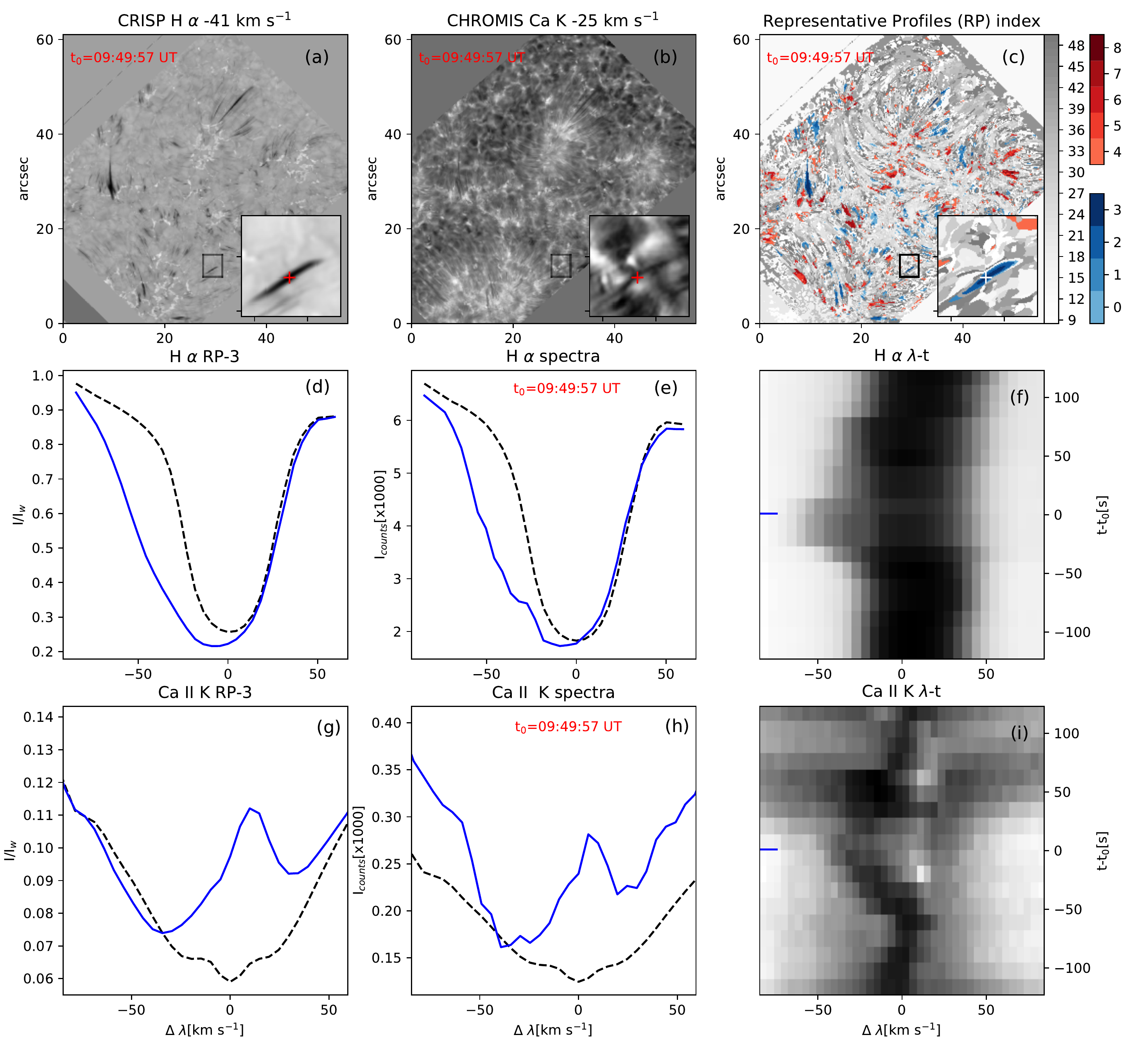}
   \caption{
   Identification of RBEs in \halpha\ and \cak. The top row shows blue wing \halpha\ ($a$) and \cak\ ($b$) spectroheliograms, and the same region color-coded in terms of representative profile (RP) indexing ($c$). Small insets zoom in on a selected RBE with a cross marking the location for the profiles shown in blue in ($e$) and ($h$), and the spectral evolution $\lambda t$-slices ($f$) and ($i$). The corresponding RP profiles are shown in ($d$) and ($g$). The blue marker in ($f$) and ($i$) indicates the time t$_{0}$=09:49:57 UTC. Reference quiet Sun profiles (RP-43) are shown as dashed lines. 
   } 
        \label{figure:RBES_kmeans}%
    \end{figure*}

Spicules are ubiquitous and highly dynamic features that permeate the solar
chromosphere. 
They can be divided into two categories, of which the second class (aka type-II spicules) are more dynamic with shorter lifetimes, vigorous sideways motion, and high apparent velocities \citep[][]{Bart_2007_PASJ,Tsiropoula_2012,Tiago_2012,2016ApJ...824...65P}.
Type-II spicules can be heated beyond chromospheric temperatures and during their lifetime become visible in transition region \citep{Bart_2011_Science,Tiago_2014_heat,Luc_2015}
and coronal diagnostics \citep{Bart_2011_Science,Vasco_2016,2016ApJ...830..133K,2017ApJ...845L..18D}. 
These characteristics form the basis for attributing type-II spicules an important role in mass-loading and heating of the solar corona
\citep[see, e.g.,][]{Juan_2017_Science,2017RAA....17..110T,2018SoPh..293...56K,Juan_2018} with potentially higher impact at lower temperatures \citep{2015ApJ...812L..30I}.

Type-II spicules have been observed as rapid Doppler excursions of the chromospheric \halpha\ and \ion{Ca}{ii}~8542\AA\ spectral lines, a characteristic for which they are called Rapid Blue or Red-shifted Excursions \citep[RBEs or RREs,][]{2008ApJ...679L.167L,Luc_2009,Sekse_2012,2013ApJ...769...44S,
2015ApJ...802...26K}.
The RBE (RRE) spectral signature has also been identified in \ion{Mg}{ii}~h\&k profiles \citep{Luc_2015}.

With the advent of the CHROMospheric Imaging Spectrometer (CHROMIS) at the Swedish 1-m Solar Telescope \citep[SST,][]{2003SPIE.4853..341S}, it is now possible to achieve imaging spectroscopy at an unprecedented spatial resolution of better than 0\farcs1 at wavelengths shortward of 4000\AA. 
Through high spectral resolution imaging in the \cak{}~3934\AA\ line, CHROMIS unlocks the potential to uncover spicule properties at ever smaller spatial and temporal scales.

The \cak{} spectral line features similar formation properties as \Mgk{} showing wide damping wings and self reversals in the respective line cores. 
The \cak{} line is formed lower in the atmosphere than its \element{Mg} counterpart, mostly due to roughly 18 times lower abundance of calcium compared with magnesium
\citep[for recent studies of \ion{Mg}{ii} and \ion{Ca}{ii} spectral line formation, see][]{2013ApJ...772...90L, 
2018A&A...611A..62B}.

In this letter, we use coordinated on-disk observations from the SST and Interface Region Imaging Spectrograph \citep[IRIS,][]{Bart2014}.
Type-II spicules are characterized, for the first time, in \cak{}, with the help of an unsupervised machine learning algorithm that provides a robust characterization of millions of profiles through representative bins, with only a handful being spicules. We exploit the co-temporal and co-spatial data-sets from IRIS to identify corresponding \Mgk{} spicule spectra. 
We further our understanding of spicules and spicule line-profile formation through a simple numerical experiment.


\section{Observations and Methods}
\label{Section:OBS}

We observed an enhanced network region close to disk center on 25 May 2017 for 97~min. 
From the SST, we analyzed imaging spectroscopy data in \halpha\ acquired with the CRisp Imaging SpectroPolarimeter \citep[CRISP,][]{Crisp_2008} and \cak\ with CHROMIS.
From IRIS, we analyzed concurrent \Mgk\ spectra from a 2\farcs32~$\times$~69\farcs2 raster and 2796\AA\ Slit-Jaw Images (SJIs).
The IRIS and CRISP datasets were co-aligned, and blown up to the CHROMIS pixel scale (0\farcs037) by cross-correlating \cak{} inner line-wing with SJI~2796\AA, and the photospheric wideband channels associated with the \cak{} and \halpha{} data, respectively.
For more details on the observations and data processing, we refer to appendix~\ref{Append:obs}.

Identifying rapid excursions in \ion{Ca}{ii}~K is not straightforward as they neither have a sharp image contrast as in \halpha{}
(see Fig.~\ref{figure:RBES_kmeans}$a$ and $b$), 
nor a simple absorption profile. Therefore, we use the robust \textit{k-means clustering} technique to identify RBEs and RREs in both these spectral regions simultaneously, based on their profiles as described in appendix \ref{Append:k-mean}.

\begin{figure*}
   \centering
   \includegraphics[width=\textwidth]{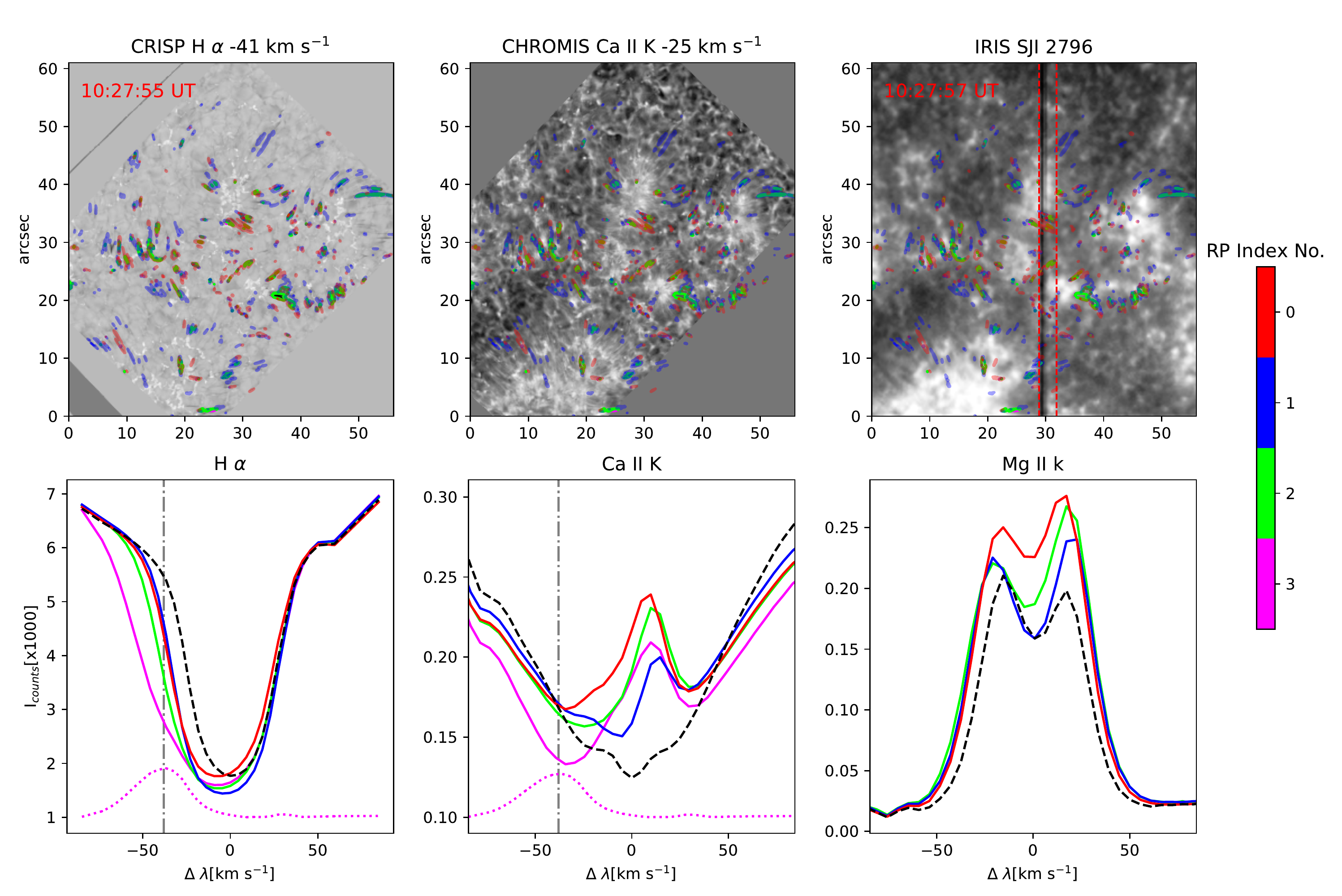}
   \caption{RBEs in \halpha, \cak\ and \Mgk. Top: spectroheliograms and slit-jaw as labelled with overplotted color-coded RP bins. Bottom: spectra for the RPs following the same color code. The dotted pink profile is the difference between the average H-alpha profile (dashed) and that of RP-3 in arbitrary units. The vertical dashed-dotted line highlights the proximity of the latter with the minimum of the \ion{Ca}{ii}~K$_3$ RP-3 at $-$38.6 \kms{}. The dashed red lines in the SJI indicates the extent of the IRIS rasters.
   }
    \label{figure:RBEs_Mg}%
    \end{figure*}
    
\begin{figure*}
   \centering
   \includegraphics[width=\textwidth]{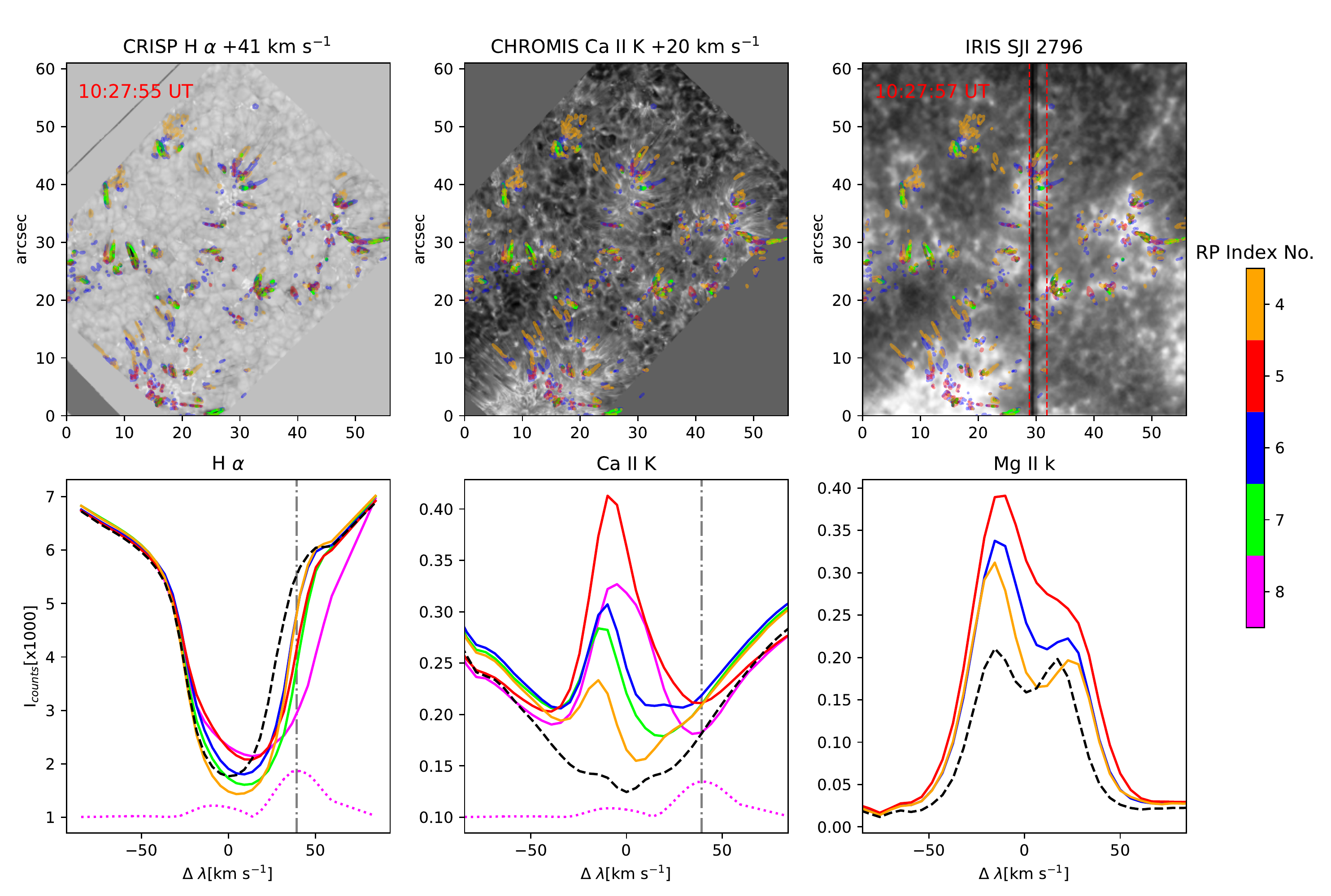}
   \caption{RREs in \halpha, \cak\ and \Mgk\ in the same format as Fig.~\ref{figure:RBEs_Mg}.
   }
    \label{figure:RREs_Mg}%
    \end{figure*}

\begin{figure}
   \centering
   \includegraphics[width=0.49\textwidth]{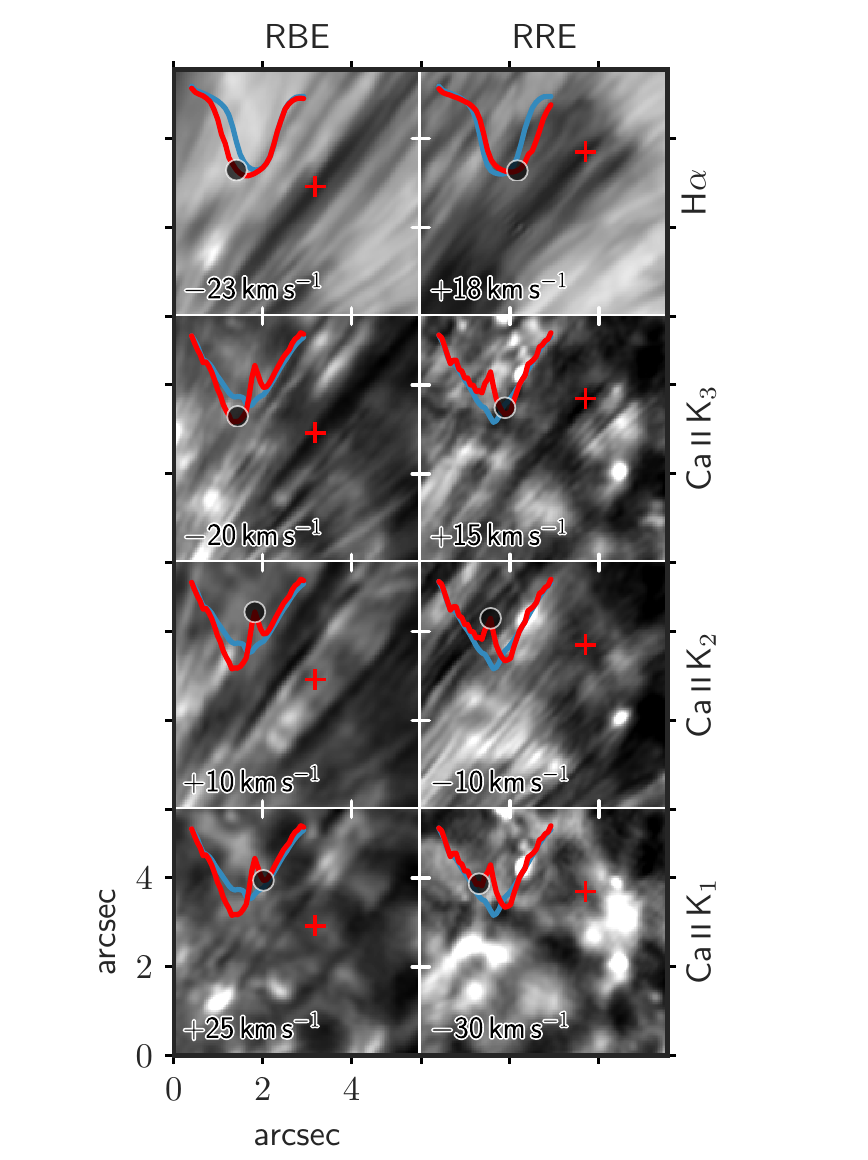}
   \caption{
   RBE and RRE with surroundings in the \halpha\ line-wing, and their \ion{Ca}{ii}~K$_3$, K$_2$, and K$_1$ wavelengths. Crosses mark the red inset profile spatial locations and the circles mark the selected wavelengths. The blue profile is the median over the showed area.
   }
    \label{figure:contrast}%
\end{figure}

\begin{figure*}
   \centering
   \includegraphics[width=0.95\textwidth]{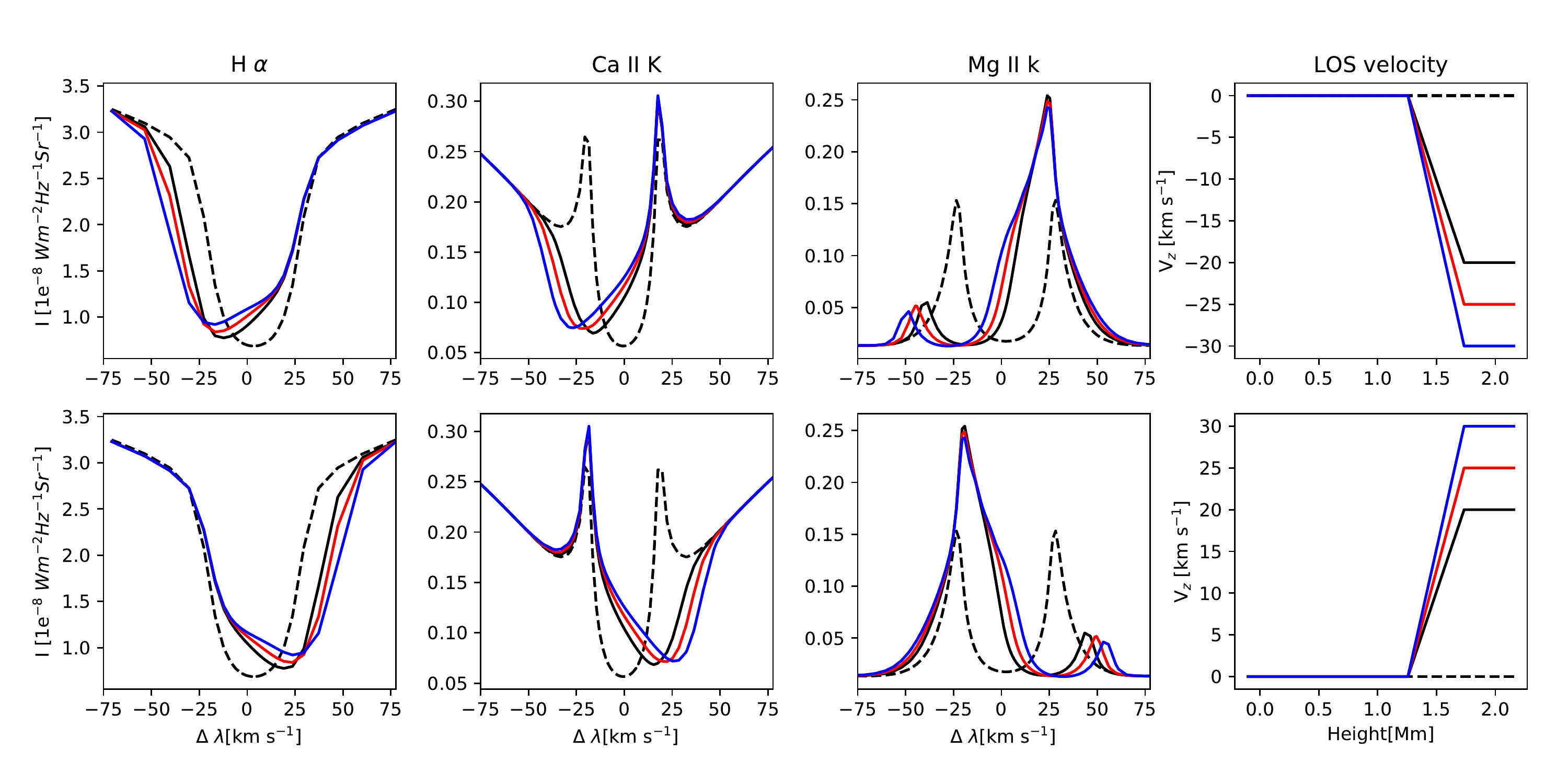}
   \caption{Forward modelling of \halpha{}, \cak{} and \Mgk{} spectra using the RH1.5D code. Top row: Synthetic RBE like profiles with a negative line-of-sight (LOS) velocity stratification with a minimum velocity of $-30$ (blue), $-25$ (red), and $-20$ (black) \kms{}. Bottom row: Synthetic RRE like profiles modelled in the same way as the RBEs except for the positive velocity stratification with a maximum velocity of $+30$ (blue), $+25$ (red), and $+20$ (black) \kms{}, respectively. The dashed black lines show the spectra with zero velocity in all the three wavelengths.
   }
    \label{figure:forward_models}%
    \end{figure*}

\section{Results and discussion}
\label{Section:results}

\subsection{Spectral characteristics of RBEs from observations and $k$-means}
\label{Section:R&D}

The $k$-means algorithm assigned each pixel on the common CRISP and CHROMIS field-of-view (FOV) to one representative profile (RP), with indices 0--49. The RP indices map corresponding to one scan is shown in Fig.~\ref{figure:RBES_kmeans}$c$. For better visibility, we have indicated RBE-like (RRE-like) RPs in shades of blue (red) and the rest of the background in gray-scale. We find that RP-0 to RP-3 show absorption in the blue wing of \halpha{} (see Fig.~\ref{figure:RBEs_Mg}) as characteristic to RBEs, and that RP-4 to RP-8 show the red-wing absorption typical of RREs (see Fig.~\ref{figure:RREs_Mg}). Darker shades of blue and red in Fig.~\ref{figure:RBES_kmeans}c imply stronger absorption in  \halpha{} wings. Consequently, RP-1 is stronger than RP-0 and so on. 

The insets in the first row of Fig.~\ref{figure:RBES_kmeans} zoom in on a dark thread-like feature that was identified as an RBE. It is evident that RBEs in blue wing \ion{Ca}{ii}~K have reduced contrast with respect to the background compared with  \halpha{}, thereby making it difficult to identify them from spectroheliograms. However, their appearance is spatially coherent in both diagnostics. The blue-colored streak in the RP map inset (Fig.~\ref{figure:RBES_kmeans}$c$) shows a clear structural resemblance to the \halpha{} RBE, thereby strengthening our confidence in the detection from the $k$-means technique.

The observed spectra and RP corresponding to the pixel indicated with the red cross mark, are shown in the second and third rows of Fig.~\ref{figure:RBES_kmeans}. The profile of interest belongs to RP-3 and is plotted in Fig.~\ref{figure:RBES_kmeans}$d$ for \halpha. This profile displays significant asymmetry in the blue wing as compared to the reference quiet Sun (RP-43, dashed line), confirmed by the observed \halpha{} spectrum shown in panel $e$. The temporal evolution of the spectra is shown in the wavelength-time ($\lambda$t-slice) (Fig.~\ref{figure:RBES_kmeans}$f$) confirming the RBE "rapidity" by clearly showing its short lifetime (of \textasciitilde  50 s in this case) and the characteristic blue-ward asymmetry. We also show the RP, observed spectra and $\lambda$t-slice for the corresponding \ion{Ca}{ii}~K RBE in the bottom row of Fig.~\ref{figure:RBES_kmeans}. More than 90~\% of the events detected using $k$-means have a lifetime <~100 s. The detailed statistical analysis shall be presented in a forthcoming article.  Both the representative and observed profiles have higher specific intensity compared to the quiet Sun and are significantly asymmetric with an enhanced K$_{\textrm{2R}}$ peak
\citep[we follow the classic nomenclature for the K$_1$ to K$_3$ spectral features in the \cak\ line, see, e.g., Fig.~1 in][]{Rutten_H&K}.

The \ion{Ca}{ii}~K $\lambda$t-slice shows a clearer progression of the rapid blue-ward excursion, due to the higher cadence and narrower line-minimum compared with \halpha{}, and reveals continuity between the line-minimum at rest and the line-minimum at maximum shift. Thus, the line-minimums for the most strongly-shifted spicules are likely the usual K$_3$ line features, merely Doppler shifted. The presented RPs in Fig.~\ref{figure:RBES_kmeans} are normalized as used for the $k$-means (see appendix~\ref{Append:k-mean}) while the observed spectral pair of \halpha{} and \cak{} have been plotted in un-normalized intensity to compare the absolute differences between the RBE and the quiet background.

From Fig.~\ref{figure:RBEs_Mg} and Fig.~\ref{figure:RREs_Mg}, we see that the strongest Doppler shifts of the \cak{} line-minimum, those of RP-0, 5, and 8, are associated with the strongest K$_2$ features at the opposite wavelengths. The line-minimum of RP-3 has the same Doppler shift as RP-0 but approximately the same K$_2$ intensity when measured relatively to its line-minimum. RP-1 and 4 show both the smallest shifts and the weakest K$_2$ peaks. Together with the completeness of the $k$-means clustering method and the reduced number of clusters where spicules are present, this establishes a clear relation between the Doppler-shift of the line-minimum and the brightness of the K$_2$ peaks.

Besides intensity, the K$_2$ features seem to strengthen in width, but the different peaks differ the most in the wavelengths towards the line-minimum. Notably, for RBEs, there is no widening of the K$_2$ in the direction opposite to the new line-minimum, with the associated K$_1$ and immediate surrounding line-wing slopes being almost identical for all RPs. For RREs the widening is overwhelmingly towards the shifted K$_3$; thus, the width change of the K$_2$ features is likely associated with the removal of the K$_3$ feature from the wavelengths where the widening is observed.

The maxima of the \halpha{} differential profiles \citep[as in][]{Luc_2009} for RP-3 and 8, plotted in Figs.~\ref{figure:RBEs_Mg} and \ref{figure:RREs_Mg}, nearly match the shifted \cak{} line-minima. Such match is indicative of a physical meaning of such wavelength position.

Motivated partially by a similar asymmetric broadening of \ion{Ca}{ii}~K$_2$, \citet{1970SoPh...11..347A} modeled K$_{2}$ grain enhancements using strong flows of opposite direction to that expected by their spectral position and found differential opacity shifts dominating the enhancements, with flow-driven source function changes being small and relatively unimportant. Flow-driven source function enhancements can be important for K$_3$ formation \citep{1981ApJ...249..720S,Scharmer_1984}, via capture of deeper more intense radiation. With time-dependent forward modeling, \citet{Carlsson_1997} explained the formation of \ion{Ca}{ii}~H\&K bright grains, where velocity gradients were important in amplifying shock-driven source function enhancements. Similarly, \cite{2015ApJ...810..145D} found a non-LTE source function increase of the \ion{Ca}{ii}~8542 line, enhanced in the red and suppressed in the blue via a differential upflow.
In this trend, non-LTE \ion{Ca}{ii}~8542 inversions by \citet{Vasco2017-Flashes} and \citet{Bose_2019} showed that strong differential downflows in the upper atmosphere would best generate the blue-shifted emission features of umbral flashes, with \citet{Bose_2019} finding evidence that such downflows could explain observed asymmetries in the \Mgk{} profiles.

Based on the relations identified in this discussion, we propose that the line formation of the most strongly shifted RBEs (RREs) is dominated by a combination of the Doppler shift of the K$_3$, suppressing completely one of the K$_2$ peaks, and an "opacity window" effect, where K$_2$ features opposite in wavelength to the flow direction are enhanced "in-place" due to differential Doppler removal of upper-layer opacity. For the RBE case the K$_2$ suppression is such that their intensity is as low as that of the quiet average. Such insensitivity to the lower layers, together with the match with the \halpha{} differential profiles, further indicates that the line-minimum is the Doppler shifted K$_3$, and that its position is representative of the true mass velocity of the spicule, being the defining feature to search for in spicule \cak{} studies. 

While symmetrical in shape to RBEs and featuring the same relations that lead us to the identification of both the opacity window effect and of a flow-dominated K$_3$, RRE RPs are different from RBE RPs in intensity levels. For both \Mgk{} and \cak{} the overall light levels of most RRE RPs depart significantly from the quiet reference case and show elevated line-minimums in \halpha. RBEs do not. This suggests the contribution of other line-forming mechanisms to RREs, possibly related to heating (but then leading to raised \halpha{} profiles rather than mere broadening), in a way beyond the scope of this letter.

\subsection{Where do we "see" the RBEs and RREs in \cak{}?}
\label{Section:Seespicules}
From the spectroheliograms observed in the \cak{} it is clear that, unlike \halpha{}, identifying spicules based on their contrast with respect to the background is a challenging task. We selected one RBE and one RRE, in \halpha{} and \cak{}, as shown in Fig.~\ref{figure:contrast}, with more examples shown in appendix~\ref{Append:RBE_RRE_examples}. From the former, it is clear that the same spicules that are present in \halpha\ (top row) are also present, with almost the same shape and absorption character, in the line-minimum of \cak{} (second row) which, as visible in the inset profiles, is the Doppler-shifted K$_{3}$. In the third row of Fig.~\ref{figure:contrast}, where the K$_2$ wavelengths have been selected to generate an image, the contrast of the RBEs and RREs is visibly much lower than the contrast of the K$_{3}$ feature. More strikingly, the patterns such as bright points and bright lanes from the lower layers (bottom row) are distinctly visible in the K$_{2}$ image, whereas for the K$_{3}$ images no background patterns are ever discernible along the body of the spicule. This is consistent with the line formation hypothesis put forward, i.e., when we observe the K$_{2}$ feature of RBEs (RREs) (as we do in the third row of Fig.~\ref{figure:contrast}), we primarily observe a background-driven enhancement via photons shining through the "window".

\subsection{RBEs and RREs in \Mgk{}}
 
\label{Section:RBES_Mg}

Figures~\ref{figure:RBEs_Mg} and \ref{figure:RREs_Mg} include co-temporal and co-spatial SST and IRIS SJIs and spectra, matching the colour scheme used for \cak{}. The core of the \Mgk{} spectra are Doppler shifted to the blue (red) in tandem with the shifts in \cak{} core and at the same time, the opposite k$_{2}$ peaks are enhanced, suggesting the same opacity shift and window effects present as in \cak{}.
Such effects do not seem as pronounced as for the \cak{} spectra. This may be due to the slenderness of the spicules, leading to mixing of profiles for the lower resolution IRIS data, further exacerbated by the two-pixel binning. Moreover, \Mgk{} spectra corresponding to the RPs with the strongest k$_3$ absorption are absent, as such spicules occurred outside of the IRIS slit sampling. Still, from the at-rest case (black dashed profile) to all RBE and RRE RPs, one sees the expected relationship changes, with a k$_3$ shift leading to a suppressed k$_\mathrm{2}$ and opposite enhanced k$_\mathrm{2}$, matching their corresponding \cak{} relations. Note that RP-0 k$_{3}$ has nearly the same Doppler shift as that of RP-1, and likewise nearly the same \halpha\ asymmetry. Unsurprisingly, they show a similar k$_{2}$ to k$_{3}$ relation. 

Both the RBE and the RRE \Mgk{} profiles have a higher intensity compared to the average background. Further, the RP-5 in Fig.~\ref{figure:RREs_Mg}, the most strongly shifted in \cak{} that is present in \Mgk{}, shows a complete suppression of k$_{\mathrm{2R}}$. A very similar complete suppression, but for the RBE case, can be seen in Fig.1$g$ of \citet{Luc_2015} which, in turn, is at the maximum excursion of the k$_3$ ($\lambda t$ diagram  therein). This literature example fits, and completes, the relations between spectral features that we identify for both \cak{} and \Mgk{}, and is qualitatively reproduced by the following numerical experiment. 

\subsection{Radiative transfer computations}

\label{Section:Rad_trans}

We performed a simple forward modelling based on radiative transfer computations where all the three spectral profiles, \halpha{}, \cak{} and \Mgk{} were synthesized using the \verb|RH1.5D| code \citep{Uitenbroek_2001,Tiago_RH_2015}. We used the standard 5 level plus continuum hydrogen (\ion{H}{i}) and calcium (\ion{Ca}{ii}) models from \verb|RH1.5D| and a 10 level plus continuum model of \ion{Mg}{ii}. The profiles were synthesized using the \verb|FAL-C| atmosphere \citep{FALC1993} with modified LOS velocity stratifications as shown in Fig.~\ref{figure:forward_models}. These consisted of a gradient in velocity (Fig.~\ref{figure:forward_models}, rightmost column), at a height where the core of \cak{} is most sensitive, with maximum velocities of $\pm30$, $\pm25$ and $\pm20$~\kms{}.

The two rows of Fig.~\ref{figure:forward_models} shows the synthetic RBE (RRE)-like intensity profiles for three different negative (positive) LOS velocity stratifications. The dashed spectra plot the at-rest case. 

In this simple experiment, we see that, as apparent from the observations, the Doppler-shift of the K$_{3}$ and k$_{3}$ features are the same as the velocities of the upper layers; that they suppress a K$_{2}$ and k$_{2}$ peak, and that the opposite-in-offset K$_{2}$ and k$_{2}$ peaks are enhanced without a Doppler-shift, widened in the direction of the line-core as would be explained by the "window" effect.

The main difference between the \cak{} and the \Mgk{} cases is that the latter shows a residual peak that could be erroneously interpreted as a flow at that Doppler-shift.


\section{Conclusions}

We find that \cak{} and \Mgk{} spicule line formation is dominated by opacity shifts and associated opacity "windows". We find indications for other mechanisms, likely heating-based, in the enhancement of the K$_\mathrm{2V}$ (k$_\mathrm{2V}$) peaks for RREs but opacity-shift and "window" relations are always present and dominant. 

We find that a basic model, generated only with this understanding and a simple gradient in velocity from at-rest to strong velocities in the upper chromospheric layers, effectively reproduces the main properties of the observed spicule profiles. .

The wavelength position of the K$_2$ (k$_2$) peaks is not directly usable as a velocity diagnostic for spicules. In practical terms, the Doppler shift of the K$_{3}$ (k$_{3}$) features provides an accurate velocity measure, one consistent across multiple lines.
The spicule Doppler shifts that we measure from \cak\ to be in the range $20-50$~\kms\ confirm earlier spicule Doppler measurements from other spectral lines \citep{Luc_2009,2016ApJ...830..133K}, largely removing concern over statistics based on \halpha\ differential profiles and reservations about the presence of such flows \citep{2011ApJ...730L...4J}. These Doppler measurements constitute the LOS component of the mass flow in spicules which, due to our top-down viewing angle, can be directly compared to the apparent spicule velocities, measured nearly perpendicularly to the limb, of up to 150~\kms\ \citep{Bart_2007_PASJ,Tiago_2012,Tiago_2014_heat,2015ApJ...806..170S}, if those apparent velocities were due to mass flows. Due to the completeness of our $k$-mean clustering analysis and the absence of such high Doppler velocities, this is resolving evidence that >100~\kms\ limb apparent motions cannot be attributed to mass flows and must be due to other mechanisms such as rapidly propagating heating fronts \citep{2017ApJ...849L...7D}.

\begin{acknowledgements}
We thank Ainar Drews for his help with the observations and Rob Rutten, Tiago Pereira and Mats Carlsson for their suggestions.
\verb|RH1.5D| is publicly available at \url{https://github.com/ITA-Solar/rh}. IRIS is a NASA small explorer mission developed and operated by LMSAL with mission operations executed at the NASA Ames Research center and major contributions to downlink communications funded by ESA and the Norwegian Space Centre. The Swedish 1-m Solar Telescope is operated on the island of La Palma by the Institute for Solar Physics of Stockholm University in the Spanish Observatorio del Roque de los Muchachos of the Instituto de Astrof\'{i}sica de Canarias. The Institute for Solar Physics is supported by a grant for research infrastructures of national importance from the Swedish Research Council (registration number 2017-00625). This research is supported by the Research Council of Norway, project number 250810, and through its Centers of Excellence scheme, project number 262622.
\end{acknowledgements}

\bibliographystyle{aa} 
\bibliography{references1.bib} 

\begin{thebibliography}{47}
\expandafter\ifx\csname natexlab\endcsname\relax\def\natexlab#1{#1}\fi

\bibitem[{Arthur \& Vassilvitskii(2007)}]{arthur2007k}
Arthur, D. \& Vassilvitskii, S. 2007, in Proceedings of the eighteenth annual
  ACM-SIAM symposium on Discrete algorithms, Society for Industrial and Applied
  Mathematics, 1027--1035

\bibitem[{{Athay}(1970)}]{1970SoPh...11..347A}
{Athay}, R.~G. 1970, \solphys, 11, 347

\bibitem[{{Bj{\o}rgen} {et~al.}(2018){Bj{\o}rgen}, {Sukhorukov}, {Leenaarts},
  {Carlsson}, {de la Cruz Rodr{\'\i}guez}, {Scharmer}, \&
  {Hansteen}}]{2018A&A...611A..62B}
{Bj{\o}rgen}, J.~P., {Sukhorukov}, A.~V., {Leenaarts}, J., {et~al.} 2018, \aap,
  611, A62

\bibitem[{{Bose} {et~al.}(2019){Bose}, {Henriques}, {Rouppe van der Voort}, \&
  {Pereira}}]{Bose_2019}
{Bose}, S., {Henriques}, V. M.~J., {Rouppe van der Voort}, L., \& {Pereira}, T.
  M.~D. 2019, \aap, 627, A46

\bibitem[{{Carlsson} \& {Stein}(1997)}]{Carlsson_1997}
{Carlsson}, M. \& {Stein}, R.~F. 1997, \apj, 481, 500

\bibitem[{{de la Cruz Rodr{\'\i}guez} {et~al.}(2015{\natexlab{a}}){de la Cruz
  Rodr{\'\i}guez}, {Hansteen}, {Bellot-Rubio}, \&
  {Ortiz}}]{2015ApJ...810..145D}
{de la Cruz Rodr{\'\i}guez}, J., {Hansteen}, V., {Bellot-Rubio}, L., \&
  {Ortiz}, A. 2015{\natexlab{a}}, \apj, 810, 145

\bibitem[{{de la Cruz Rodr{\'\i}guez} {et~al.}(2015{\natexlab{b}}){de la Cruz
  Rodr{\'\i}guez}, {L{\"o}fdahl}, {S{\"u}tterlin}, {Hillberg}, \& {Rouppe van
  der Voort}}]{2015A&A...573A..40D}
{de la Cruz Rodr{\'\i}guez}, J., {L{\"o}fdahl}, M.~G., {S{\"u}tterlin}, P.,
  {Hillberg}, T., \& {Rouppe van der Voort}, L. 2015{\natexlab{b}}, \aap, 573,
  A40

\bibitem[{{De Pontieu} {et~al.}(2017{\natexlab{a}}){De Pontieu}, {De Moortel},
  {Martinez-Sykora}, \& {McIntosh}}]{2017ApJ...845L..18D}
{De Pontieu}, B., {De Moortel}, I., {Martinez-Sykora}, J., \& {McIntosh}, S.~W.
  2017{\natexlab{a}}, \apjl, 845, L18

\bibitem[{{De Pontieu} {et~al.}(2017{\natexlab{b}}){De Pontieu},
  {Mart{\'{\i}}nez-Sykora}, \& {Chintzoglou}}]{2017ApJ...849L...7D}
{De Pontieu}, B., {Mart{\'{\i}}nez-Sykora}, J., \& {Chintzoglou}, G.
  2017{\natexlab{b}}, \apjl, 849, L7

\bibitem[{{De Pontieu} {et~al.}(2007){De Pontieu}, {McIntosh}, {Hansteen},
  {Carlsson}, {Schrijver}, {Tarbell}, {Title}, {Shine}, {Suematsu}, \&
  {Tsuneta}}]{Bart_2007_PASJ}
{De Pontieu}, B., {McIntosh}, S., {Hansteen}, V.~H., {et~al.} 2007, \pasj, 59,
  S655

\bibitem[{{De Pontieu} {et~al.}(2011){De Pontieu}, {McIntosh}, {Carlsson},
  {Hansteen}, {Tarbell}, {Boerner}, {Martinez-Sykora}, {Schrijver}, \&
  {Title}}]{Bart_2011_Science}
{De Pontieu}, B., {McIntosh}, S.~W., {Carlsson}, M., {et~al.} 2011, Science,
  331, 55

\bibitem[{{De Pontieu} {et~al.}(2014){De Pontieu}, {Title}, {Lemen}, {Kushner},
  {Akin}, {Allard}, {Berger}, {Boerner}, {Cheung}, {Chou}, {Drake}, {Duncan},
  {Freeland}, {Heyman}, {Hoffman}, {Hurlburt}, {Lindgren}, {Mathur}, {Rehse},
  {Sabolish}, {Seguin}, {Schrijver}, {Tarbell}, {W{\"u}lser}, {Wolfson},
  {Yanari}, {Mudge}, {Nguyen-Phuc}, {Timmons}, {van Bezooijen}, {Weingrod},
  {Brookner}, {Butcher}, {Dougherty}, {Eder}, {Knagenhjelm}, {Larsen},
  {Mansir}, {Phan}, {Boyle}, {Cheimets}, {DeLuca}, {Golub}, {Gates}, {Hertz},
  {McKillop}, {Park}, {Perry}, {Podgorski}, {Reeves}, {Saar}, {Testa}, {Tian},
  {Weber}, {Dunn}, {Eccles}, {Jaeggli}, {Kankelborg}, {Mashburn}, {Pust},
  {Springer}, {Carvalho}, {Kleint}, {Marmie}, {Mazmanian}, {Pereira}, {Sawyer},
  {Strong}, {Worden}, {Carlsson}, {Hansteen}, {Leenaarts}, {Wiesmann},
  {Aloise}, {Chu}, {Bush}, {Scherrer}, {Brekke}, {Martinez-Sykora}, {Lites},
  {McIntosh}, {Uitenbroek}, {Okamoto}, {Gummin}, {Auker}, {Jerram}, {Pool}, \&
  {Waltham}}]{Bart2014}
{De Pontieu}, B., {Title}, A.~M., {Lemen}, J.~R., {et~al.} 2014, \solphys, 289,
  2733

\bibitem[{Everitt(1972)}]{everitt_1972}
Everitt, B.~S. 1972, British Journal of Psychiatry, 120, 143–145

\bibitem[{{Fontenla} {et~al.}(1993){Fontenla}, {Avrett}, \&
  {Loeser}}]{FALC1993}
{Fontenla}, J.~M., {Avrett}, E.~H., \& {Loeser}, R. 1993, \apj, 406, 319

\bibitem[{{Henriques}(2012)}]{2012A&A...548A.114H}
{Henriques}, V.~M.~J. 2012, \aap, 548, A114

\bibitem[{{Henriques} {et~al.}(2016){Henriques}, {Kuridze}, {Mathioudakis}, \&
  {Keenan}}]{Vasco_2016}
{Henriques}, V.~M.~J., {Kuridze}, D., {Mathioudakis}, M., \& {Keenan}, F.~P.
  2016, \apj, 820, 124

\bibitem[{{Henriques} {et~al.}(2017){Henriques}, {Mathioudakis},
  {Socas-Navarro}, \& {de la Cruz Rodr{\'{\i}}guez}}]{Vasco2017-Flashes}
{Henriques}, V.~M.~J., {Mathioudakis}, M., {Socas-Navarro}, H., \& {de la Cruz
  Rodr{\'{\i}}guez}, J. 2017, \apj, 845, 102

\bibitem[{{Iijima} \& {Yokoyama}(2015)}]{2015ApJ...812L..30I}
{Iijima}, H. \& {Yokoyama}, T. 2015, \apjl, 812, L30

\bibitem[{{Judge} {et~al.}(2011){Judge}, {Tritschler}, \& {Chye
  Low}}]{2011ApJ...730L...4J}
{Judge}, P.~G., {Tritschler}, A., \& {Chye Low}, B. 2011, \apjl, 730, L4

\bibitem[{{Kontogiannis} {et~al.}(2018){Kontogiannis}, {Gontikakis},
  {Tsiropoula}, \& {Tziotziou}}]{2018SoPh..293...56K}
{Kontogiannis}, I., {Gontikakis}, C., {Tsiropoula}, G., \& {Tziotziou}, K.
  2018, \solphys, 293, 56

\bibitem[{{Kuridze} {et~al.}(2015){Kuridze}, {Henriques}, {Mathioudakis},
  {Erd{\'e}lyi}, {Zaqarashvili}, {Shelyag}, {Keys}, \&
  {Keenan}}]{2015ApJ...802...26K}
{Kuridze}, D., {Henriques}, V., {Mathioudakis}, M., {et~al.} 2015, \apj, 802,
  26

\bibitem[{{Kuridze} {et~al.}(2016){Kuridze}, {Zaqarashvili}, {Henriques},
  {Mathioudakis}, {Keenan}, \& {Hanslmeier}}]{2016ApJ...830..133K}
{Kuridze}, D., {Zaqarashvili}, T.~V., {Henriques}, V., {et~al.} 2016, \apj,
  830, 133

\bibitem[{{Langangen} {et~al.}(2008){Langangen}, {De Pontieu}, {Carlsson},
  {Hansteen}, {Cauzzi}, \& {Reardon}}]{2008ApJ...679L.167L}
{Langangen}, {\O}., {De Pontieu}, B., {Carlsson}, M., {et~al.} 2008, \apjl,
  679, L167

\bibitem[{{Leenaarts} {et~al.}(2013){Leenaarts}, {Pereira}, {Carlsson},
  {Uitenbroek}, \& {De Pontieu}}]{2013ApJ...772...90L}
{Leenaarts}, J., {Pereira}, T.~M.~D., {Carlsson}, M., {Uitenbroek}, H., \& {De
  Pontieu}, B. 2013, \apj, 772, 90

\bibitem[{{L{\"o}fdahl} {et~al.}(2018){L{\"o}fdahl}, {Hillberg}, {de la Cruz
  Rodriguez}, {Vissers}, {Scharmer}, {Hagfors Haugan}, \&
  {Fredvik}}]{2018arXiv180403030L}
{L{\"o}fdahl}, M.~G., {Hillberg}, T., {de la Cruz Rodriguez}, J., {et~al.}
  2018, arXiv e-prints, arXiv:1804.03030

\bibitem[{{Mart{\'\i}nez-Sykora} {et~al.}(2018){Mart{\'\i}nez-Sykora}, {De
  Pontieu}, {De Moortel}, {Hansteen}, \& {Carlsson}}]{Juan_2018}
{Mart{\'\i}nez-Sykora}, J., {De Pontieu}, B., {De Moortel}, I., {Hansteen},
  V.~H., \& {Carlsson}, M. 2018, \apj, 860, 116

\bibitem[{{Mart{\'\i}nez-Sykora} {et~al.}(2017){Mart{\'\i}nez-Sykora}, {De
  Pontieu}, {Hansteen}, {Rouppe van der Voort}, {Carlsson}, \&
  {Pereira}}]{Juan_2017_Science}
{Mart{\'\i}nez-Sykora}, J., {De Pontieu}, B., {Hansteen}, V.~H., {et~al.} 2017,
  Science, 356, 1269

\bibitem[{{Pereira} {et~al.}(2012){Pereira}, {De Pontieu}, \&
  {Carlsson}}]{Tiago_2012}
{Pereira}, T. M.~D., {De Pontieu}, B., \& {Carlsson}, M. 2012, \apj, 759, 18

\bibitem[{{Pereira} {et~al.}(2014){Pereira}, {De Pontieu}, {Carlsson},
  {Hansteen}, {Tarbell}, {Lemen}, {Title}, {Boerner}, {Hurlburt}, \&
  {W{\"u}lser}}]{Tiago_2014_heat}
{Pereira}, T.~M.~D., {De Pontieu}, B., {Carlsson}, M., {et~al.} 2014, \apj,
  792, L15

\bibitem[{{Pereira} {et~al.}(2016){Pereira}, {Rouppe van der Voort}, \&
  {Carlsson}}]{2016ApJ...824...65P}
{Pereira}, T.~M.~D., {Rouppe van der Voort}, L., \& {Carlsson}, M. 2016, \apj,
  824, 65

\bibitem[{{Pereira} \& {Uitenbroek}(2015)}]{Tiago_RH_2015}
{Pereira}, T.~M.~D. \& {Uitenbroek}, H. 2015, \aap, 574, A3

\bibitem[{{Rouppe van der Voort} {et~al.}(2015){Rouppe van der Voort}, {De
  Pontieu}, {Pereira}, {Carlsson}, \& {Hansteen}}]{Luc_2015}
{Rouppe van der Voort}, L., {De Pontieu}, B., {Pereira}, T.~M.~D., {Carlsson},
  M., \& {Hansteen}, V. 2015, \apj, 799, L3

\bibitem[{{Rouppe van der Voort} {et~al.}(2009){Rouppe van der Voort},
  {Leenaarts}, {De Pontieu}, {Carlsson}, \& {Vissers}}]{Luc_2009}
{Rouppe van der Voort}, L., {Leenaarts}, J., {De Pontieu}, B., {Carlsson}, M.,
  \& {Vissers}, G. 2009, \apj, 705, 272

\bibitem[{{Rutten} \& {Uitenbroek}(1991)}]{Rutten_H&K}
{Rutten}, R.~J. \& {Uitenbroek}, H. 1991, \solphys, 134, 15

\bibitem[{{Scharmer}(1981)}]{1981ApJ...249..720S}
{Scharmer}, G.~B. 1981, \apj, 249, 720

\bibitem[{{Scharmer}(1984)}]{Scharmer_1984}
{Scharmer}, G.~B. 1984, {Accurate solutions to non-LTE problems using
  approximate lambda operators}, ed. W.~{Kalkofen}, 173--210

\bibitem[{{Scharmer} {et~al.}(2003{\natexlab{a}}){Scharmer}, {Bjelksj{\"o}},
  {Korhonen}, {Lindberg}, \& {Petterson}}]{2003SPIE.4853..341S}
{Scharmer}, G.~B., {Bjelksj{\"o}}, K., {Korhonen}, T.~K., {Lindberg}, B., \&
  {Petterson}, B. 2003{\natexlab{a}}, in Society of Photo-Optical
  Instrumentation Engineers (SPIE) Conference Series, Vol. 4853, Innovative
  Telescopes and Instrumentation for Solar Astrophysics, ed. S.~L. {Keil} \&
  S.~V. {Avakyan}, 341--350

\bibitem[{{Scharmer} {et~al.}(2003{\natexlab{b}}){Scharmer}, {Dettori},
  {L{\"o}fdahl}, \& {Shand}}]{2003SPIE.4853..370S}
{Scharmer}, G.~B., {Dettori}, P.~M., {L{\"o}fdahl}, M.~G., \& {Shand}, M.
  2003{\natexlab{b}}, in Society of Photo-Optical Instrumentation Engineers
  (SPIE) Conference Series, Vol. 4853, Society of Photo-Optical Instrumentation
  Engineers (SPIE) Conference Series, ed. S.~L. {Keil} \& S.~V. {Avakyan},
  370--380

\bibitem[{{Scharmer} {et~al.}(2008){Scharmer}, {Narayan}, {Hillberg}, {de la
  Cruz Rodriguez}, {L{\"o}fdahl}, {Kiselman}, {S{\"u}tterlin}, {van Noort}, \&
  {Lagg}}]{Crisp_2008}
{Scharmer}, G.~B., {Narayan}, G., {Hillberg}, T., {et~al.} 2008, \apj, 689, L69

\bibitem[{{Sekse} {et~al.}(2012){Sekse}, {Rouppe van der Voort}, \& {De
  Pontieu}}]{Sekse_2012}
{Sekse}, D.~H., {Rouppe van der Voort}, L., \& {De Pontieu}, B. 2012, \apj,
  752, 108

\bibitem[{{Sekse} {et~al.}(2013){Sekse}, {Rouppe van der Voort}, {De Pontieu},
  \& {Scullion}}]{2013ApJ...769...44S}
{Sekse}, D.~H., {Rouppe van der Voort}, L., {De Pontieu}, B., \& {Scullion}, E.
  2013, \apj, 769, 44

\bibitem[{{Skogsrud} {et~al.}(2015){Skogsrud}, {Rouppe van der Voort}, {De
  Pontieu}, \& {Pereira}}]{2015ApJ...806..170S}
{Skogsrud}, H., {Rouppe van der Voort}, L., {De Pontieu}, B., \& {Pereira},
  T.~M.~D. 2015, \apj, 806, 170

\bibitem[{{Tian}(2017)}]{2017RAA....17..110T}
{Tian}, H. 2017, Research in Astronomy and Astrophysics, 17, 110

\bibitem[{{Tsiropoula} {et~al.}(2012){Tsiropoula}, {Tziotziou}, {Kontogiannis},
  {Madjarska}, {Doyle}, \& {Suematsu}}]{Tsiropoula_2012}
{Tsiropoula}, G., {Tziotziou}, K., {Kontogiannis}, I., {et~al.} 2012, \ssr,
  169, 181

\bibitem[{{Uitenbroek}(2001)}]{Uitenbroek_2001}
{Uitenbroek}, H. 2001, \apj, 557, 389

\bibitem[{{van Noort} {et~al.}(2005){van Noort}, {Rouppe van der Voort}, \&
  {L{\"o}fdahl}}]{vannoort2005MOMFBD}
{van Noort}, M., {Rouppe van der Voort}, L., \& {L{\"o}fdahl}, M.~G. 2005,
  \solphys, 228, 191

\bibitem[{{Vissers} \& {Rouppe van der Voort}(2012)}]{2012ApJ...750...22V}
{Vissers}, G. \& {Rouppe van der Voort}, L. 2012, \apj, 750, 22

\end{thebibliography}

\begin{appendix}

\section{Observations and data processing}
\label{Append:obs}

The enhanced network region was observed in a coordinated SST and IRIS campaign on 25 May 2017 and was centered on heliocentric solar coordinates $(x,y)=(31\arcsec,-89\arcsec)$ with corresponding observing angle $\mu=\cos \theta=0.99$ ($\theta$ being the heliocentric angle). The time duration of the observations was about 97 min starting from 09:12UT until 10:49UT. We used both the CRISP and CHROMIS tunable Fabry-P{\'e}rot instruments to record the data from the SST. CHROMIS was installed in 2016 and will be described in a forthcoming paper by Scharmer and collaborators. CRISP sampled the \halpha{} and \ion{Fe}{I}~6302\AA\ absorption lines in imaging spectroscopic and spectropolarimetric mode respectively at a temporal cadence of 19.6~s and spatial sampling of 0\farcs058. The \halpha{} spectra were sampled at 32 wavelength positions between $\pm$1.85~\AA\ with respect to the line core. CHROMIS sampled \cak\ at 41 wavelength positions within $\pm$1.28~\AA\ with 63.5~m\AA\ steps. In addition, a continuum position was sampled at 4000\AA. The temporal cadence of the CHROMIS data was 13.6~s with an image scale of 0\farcs038. The SST diffraction limit $\lambda/\mathrm{D}$ (with $\mathrm{D}=0.97$~m being the effective aperture) is equal to 0\farcs08 at 3934~\AA. High spatial resolution was achieved through excellent seeing conditions, the SST adaptive optics system \citep[an upgrade of the system described in][]{2003SPIE.4853..370S}, and the Multi-Object Multi-Frame Blind Deconvolution \citep[MOMFBD,][]{vannoort2005MOMFBD} image restoration technique. The CRISP pipeline \citep{2015A&A...573A..40D} and an early version of the CHROMIS pipeline \citep{2018arXiv180403030L} were used for further data reduction, both including the spectral consistency method of \cite{2012A&A...548A.114H}. The CRISP data were aligned to the CHROMIS data through cross-correlation of the photospheric wideband channels, for CRISP with a full-width at half maximum (FWHM)=4.9~\AA\ centered at the \halpha\ line and for CHROMIS with a FWHM=13.2~\AA\ centered between the \ion{Ca}{ii}~H and K lines at 3950~\AA. 

IRIS was running a so-called "medium-dense 8-step raster" program (OBS-ID 3633105426) with 2~s exposure time and continuous 0\farcs34 steps of the spectrograph slit covering a field-of-view of 2\farcs32 and 69\farcs2 in the solar-X and Y directions respectively, spatially and spectrally binned by a factor of 2. The cadence of the rasters is 25~s. In addition, IRIS recorded slit-jaw images in the 2796\AA\ channel (dominated by \Mgk{} core and inner wings) as well as 1400\AA\ with exposure times of 2~s, pixel scale of 0\farcs33 and a FOV of 64\farcs2 $\times$ 69\farcs2 in the solar X and Y directions.

The IRIS data were aligned to the SST data (blown up to the CHROMIS pixel scale) by cross-correlating the \cak{} inner line-wings with the 2796 SJI. The datasets were initially investigated using CRISPEX \citep[][]{2012ApJ...750...22V}.

\section{$k$-means clustering}
\label{Append:k-mean}

\begin{figure}
  \centering
  \includegraphics[width=0.49\textwidth]{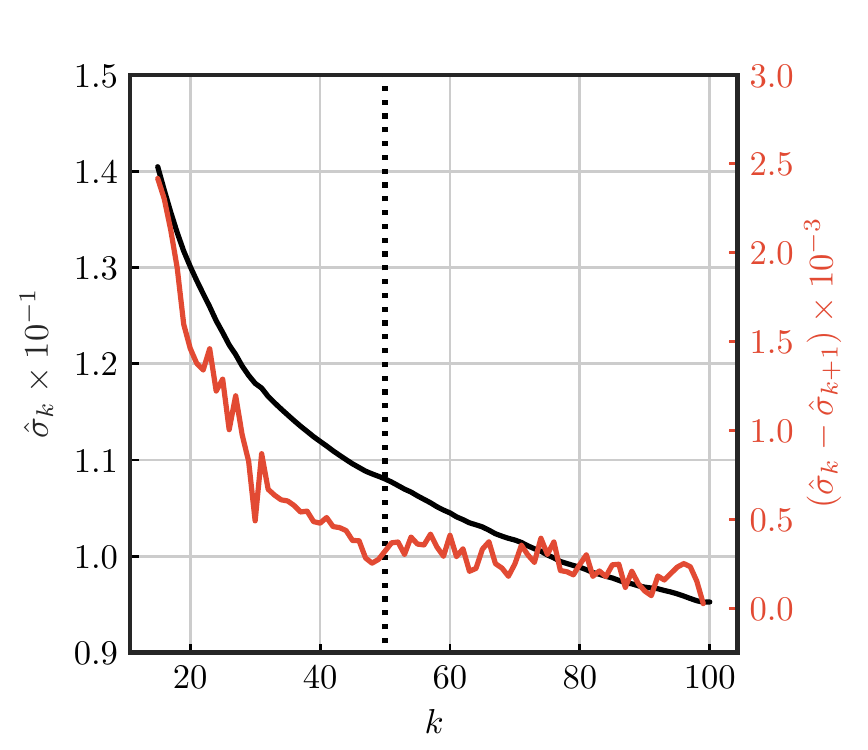}
  \caption{Finding the optimum number of clusters, $k$, for classification of the combined \halpha{} and \cak{} line profiles using the $k$-means technique. The black curve denotes, $\hat{\sigma}_{k}$, mean of standard-deviation within the clusters for \textit{k} clusters, whereas, $\hat{\sigma}_{k}-\hat{\sigma}_{k+1}$ is presented by the red curve. The estimated optimum number for clusters, $k=50$, is indicated by the dotted vertical line.}
              \label{figure:nocluster}%
\end{figure}

\begin{figure*}
  \centering
  \includegraphics[width=\textwidth]{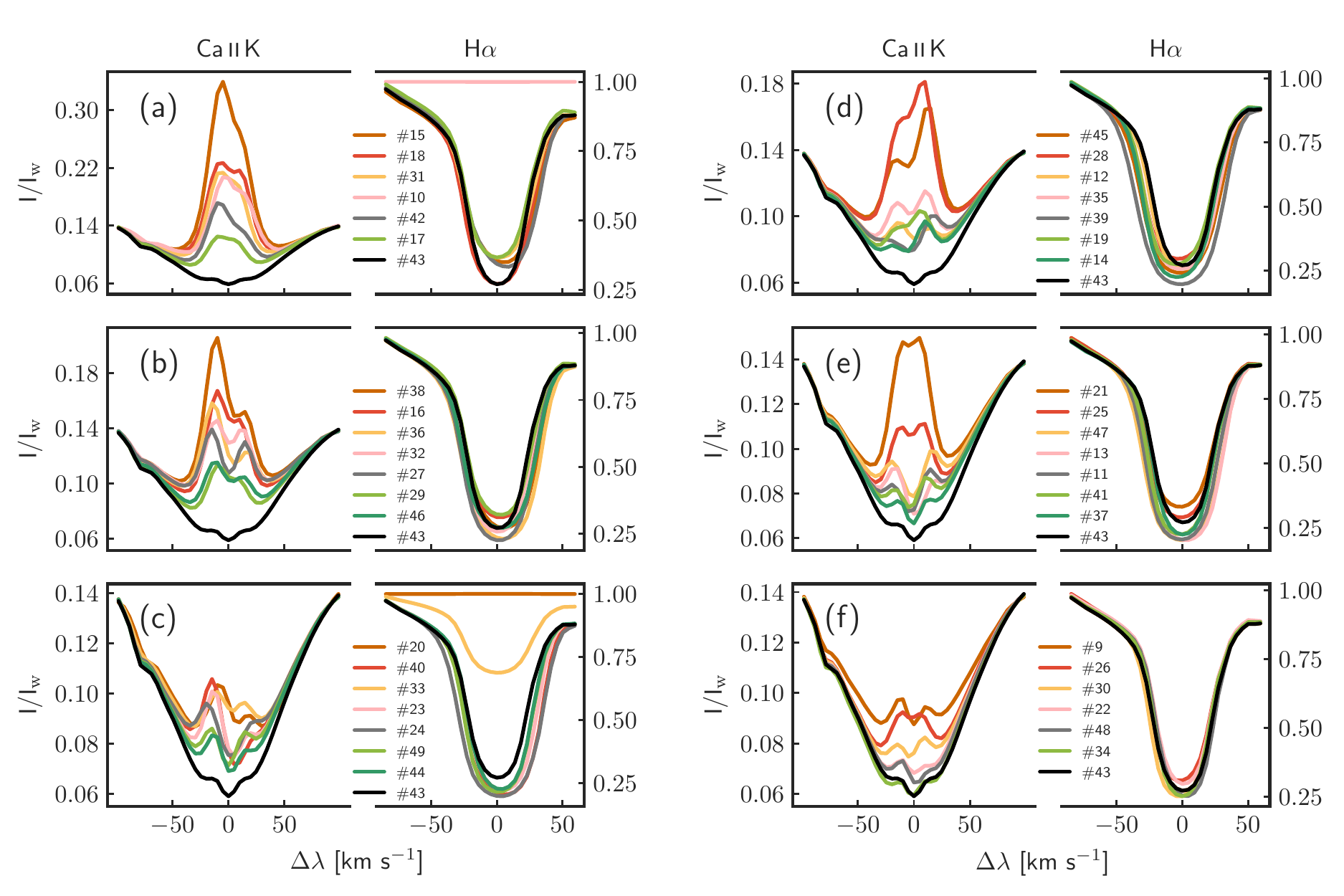}
  \caption{Representative profiles, RPs, obtained by the $k$-means classification of the \halpha{} and \cak{} line profiles. Forty one RPs (9-49) are presented in six groups in panels (a)-(f). The average quiet-Sun profile represented by RP-43 is shown in all the panels for comparison purpose. RPs 0-3 (RBE profiles) and 4-8 (RRE profiles) are plotted in Fig.~\ref{figure:RBEs_Mg} and Fig.~\ref{figure:RREs_Mg}, respectively. Here, $I_{\rm{w}}$ represents normalizing factor for the intensity, $I$.} 
              \label{figure:kmean_centers}%
    \end{figure*}

Commonly used in machine learning applications, the $k$-means clustering technique \citep[][]{everitt_1972} is defined as one where a certain number of observations in a data set is partitioned into $k$ clusters, where each observation is represented by a cluster with the closest mean. It is a very simple but robust algorithm that is primarily based on the following steps: (1) choose \textit{k} measurements as initial clusters centers; (2) compute the Euclidian distances between each observed point $\mathbf{\emph{x}}$ in the data and the cluster centers \textbf{$\mathbf{\emph{$\mu$}}$} given by $\sum_{i=1}^{n}\sum_{j=1}^{k} ||\mathbf{\emph{x}^{(i)}}-\mathbf{\emph{$\mu$}^{(j)}}||$, where $||\bullet||$ represents the Euclidian distance; (3) assign an observed point to the closest cluster center by minimizing the sum of squared errors within each cluster; (4) recompute the new cluster centers by averaging all the observations in each cluster; and (5) repeat steps (2)-(4) until none of the observations changes its cluster in two successive iterations, thereby reaching convergence. One of the major limitations is that the result of the clustering (also the number of iterations required) depends significantly on the initialization (step (1)). To circumvent this problem, we used the \verb|k-means++| \citep[][]{arthur2007k} method of initialization, where, once the first cluster center is drawn randomly, the idea is to chose the newer cluster centers in a way that they are far from the previously chosen ones. 
 
The $k$-means algorithm was used to cluster the Stokes I profiles by simultaneously combining the co-aligned \halpha{} and \ion{Ca}{ii}~K spectra for each pixel on the FOV, as shown in Fig.~\ref{figure:RBES_kmeans}. The major advantage of applying this technique to the combined spectra was to identify features based on their spectral signatures in \halpha{} and infer the corresponding signature in \cak{}.
 
Before applying the $k$-means to the observations, the \halpha{} and \cak{} spectra were normalized by a factor, $I_{\rm{w}}$. For each \halpha{} spectrum, $I_{\rm{w}}$ is the intensity of the shortest wavelength observed, whereas $I_{\rm{w}}$ for each \cak{} spectrum is also the intensity of the shortest wavelength observed but multiplied by a factor, 0.14. Here, 0.14 is the average intensity of the shortest wavelength observed in the \cak{} line over the entire FOV normalized to the average continuum intensity at 4000~\AA. This normalization is important, especially for the \cak{} spectra, to avoid large pixel-to-pixel intensity fluctuations in the line wings caused by photospheric variations, thereby allowing the algorithm to focus on the line profile between K$_{\mathrm{1v}}$ and K$_{\mathrm{1r}}$ which forms in the chromosphere. The representative profiles shown in Fig.~\ref{figure:RBES_kmeans} were normalized this way.  
 
It is important to find an optimum number of clusters to use the $k$-means classification technique more effectively. We used the mean of the standard-deviation within the clusters, $\hat{\sigma}_{k}$, to find the optimum number of clusters and for this we applied $k$-means to a sub-set of our observations (ten scans) with \textit{k} varying from 15 to 100. Variation of $\hat{\sigma}_{k}$ with respect to \textit{k} is shown Fig.~\ref{figure:nocluster}, which indicates that in general $\hat{\sigma}_{k}$ decreases with higher value of \textit{k}. In principle, $\hat{\sigma}_{k}$ should reach its minimum when \textit{k} is equal to total number of data points. However it is clear that $\hat{\sigma}_{k}$ decreases almost linearly after \textit{k} is larger than a certain value, which is $\sim50$ in our case, and that can be seen by variations in $\hat{\sigma}_{k}-\hat{\sigma}_{k+1}$ plotted in Fig.~\ref{figure:nocluster}. So, based on analysis of $\hat{\sigma}_{k}$  we choose $k=50$ for our $k$-means classification of the \halpha{} and \cak{} line profiles.

We chose ten scans (more than $2\times10^{7}$ pixels), at different time steps, from the entire time series for training the $k$-means model with 50 initial clusters. This allowed the $k$-means model to assign all the pixels of the training dataset to a particular cluster (as described above). Later, this model was applied to the entire time series that enabled an automatic and unsupervised clustering of the spectra corresponding to all the pixels in the FOV, for each and every time step. One such example is shown in the first row of  Fig.~\ref{figure:RBES_kmeans}. Each pixel is assigned to a particular cluster center that is indicated by a representative profile (RP) index. Figure~\ref{figure:kmean_centers} shows RPs for cluster number 9-49, RBEs like RPs (0-3) in the \halpha{} line and corresponding \cak{} RPs are depicted in Fig.~\ref{figure:RBEs_Mg}. Similarly, RRE like RPs (4-8) are presented in Fig.~\ref{figure:RREs_Mg}. 

We do not include the \Mgk{} line observed with IRIS in our $k$-means clustering, primarily because it has very limited spatial coverage compared to CRISP and CHROMIS. Instead, we average all  the \Mgk{} profiles spatially and temporally coinciding with a particular cluster obtained with the $k$-means clustering and this way create RPs for the \Mgk{} line which can be analyzed along with the \halpha{} and \cak{} RPs.  

\section{RBEs/RREs in \cak{} spectroheliograms}
\label{Append:RBE_RRE_examples}
 
\begin{figure}
  \centering
  \includegraphics[width=0.49\textwidth]{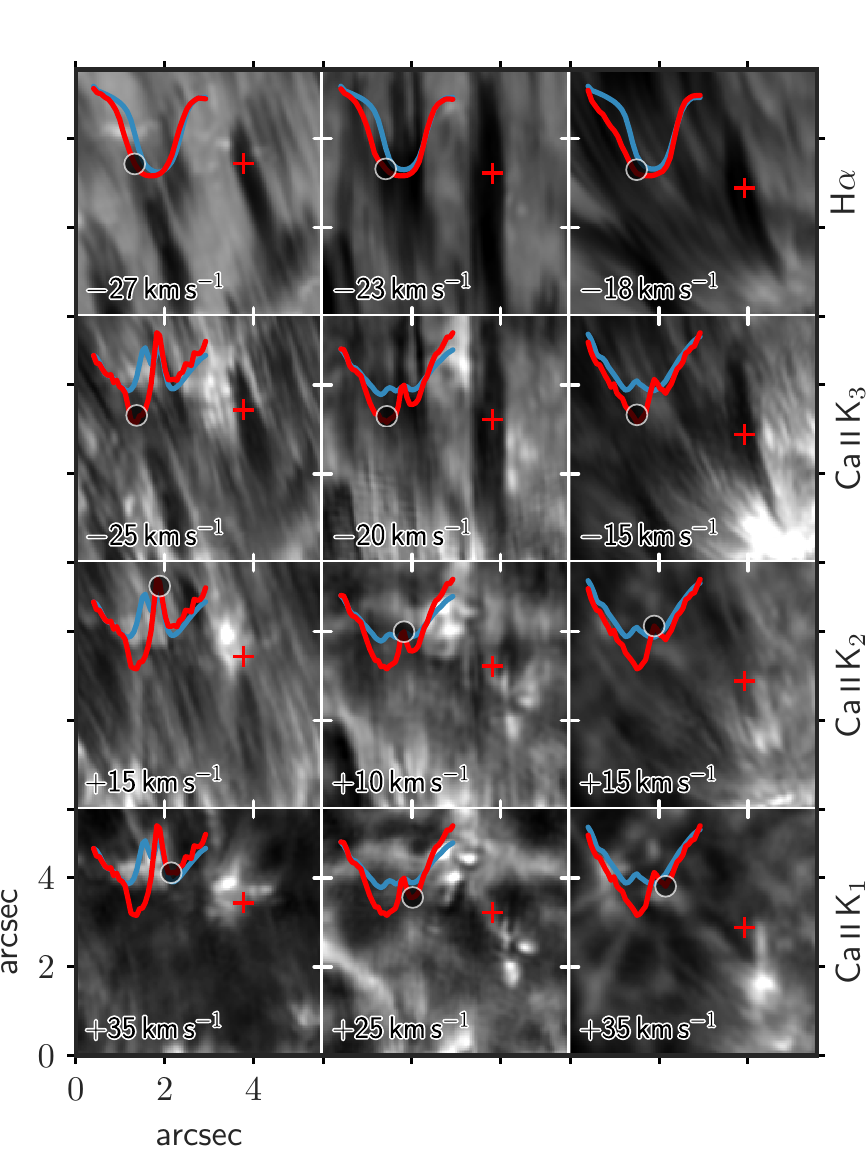}
  \caption{Three additional RBE examples in the same format as Fig.~\ref{figure:contrast}.
  }
  \label{figure:contrast_RBE}%
\end{figure}

\begin{figure}
  \centering
  \includegraphics[width=0.49\textwidth]{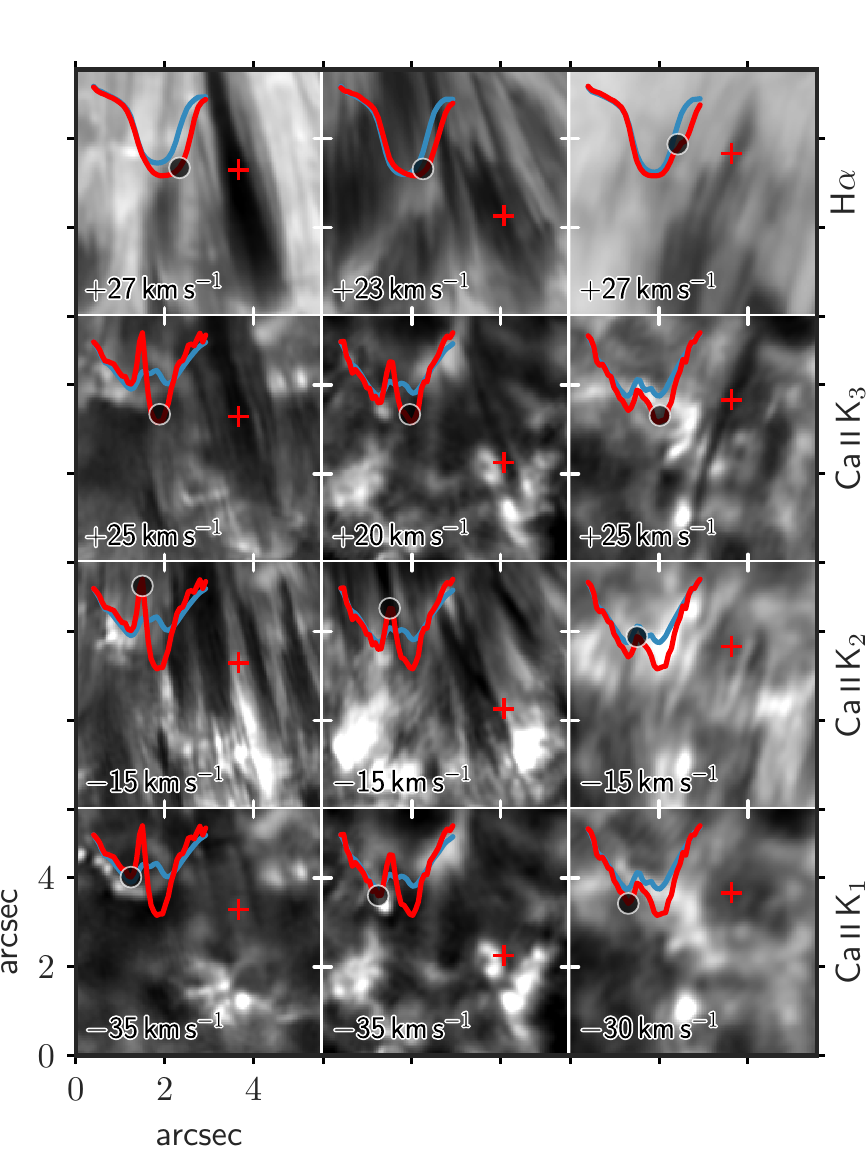}
  \caption{Three additional RRE examples in the same format as Fig.~\ref{figure:contrast}.
  }
  \label{figure:contrast_RRE}%
\end{figure}

As discussed in Sect.~\ref{Section:results}, RBEs (RREs) have Doppler shifted line minima and spectroheliograms in the corresponding \ion{Ca}{ii}~K$_3$ wavelength display a spicule morphology that is coherent to that visible in the \halpha\ wing (see Fig~\ref{figure:contrast}). The RBE (RRE) K$_2$ spectroheliogram on the other hand, is mostly reflecting the background area of the spicule. 
To demonstrate the consistency of the analysis and the interpretation put forward in Sect.~\ref{Section:Seespicules}, we present three more examples each for RBEs and RREs in Fig.~\ref{figure:contrast_RBE} and ~\ref{figure:contrast_RRE}, respectively.
 
\end{appendix}
\end{document}